# From Molecular Dynamics to hydrodynamics – a novel Galilean invariant thermostat

**Simeon D. Stoyanov and Robert D. Groot**

*Unilever Research Vlaardingen,*
*P.O. Box 114, 3130 AC Vlaardingen, The Netherlands*

**Abstract**

This article proposes a novel thermostat applicable to any particle-based dynamic simulation. Each pair of particles is thermostated either (with probability $P$) with a pairwise Lowe-Andersen thermostat [C.P. Lowe, *Europhys. Lett.* **47**, 145 (1999)], or (with probability $1–P$) with a thermostat that is introduced here, which is based on a pairwise interaction *similar* to the Nosé-Hoover thermostat. When the pairwise Nosé-Hoover thermostat dominates (low $P$), the liquid has a high diffusion coefficient and low viscosity, but when the Lowe-Andersen thermostat dominates, the diffusion coefficient is low and viscosity is high. This novel Nosé-Hoover-Lowe-Andersen thermostat is Galilean invariant and preserves both total linear and angular momentum of the system, due to the fact that the thermostatic forces between each pair of the particles are pairwise additive and central. We show by simulation that this thermostat also preserves hydrodynamics. For the (non-interacting) ideal gas at $P=0$, the diffusion coefficient diverges and viscosity is zero, while for $P>0$ it has a finite value. By adjusting probability $P$, the Schmidt number can be varied by orders of magnitude. The temperature deviation from the required value is at least an order of magnitude smaller than in Dissipative Particle Dynamics (DPD), while the equilibrium properties of the system are very well reproduced. The thermostat is easy to implement and offers a computational efficiency better than DPD, with better temperature control and greater flexibility in terms of adjusting the diffusion coefficient and viscosity of the simulated system. Applications of this thermostat include all standard molecular dynamic simulations of dense liquids and solids with any type of force field, as well as hydrodynamic simulation of multi-phase systems with largely different bulk viscosities, including surface viscosity, and of dilute gases and plasmas.

arXiv:cond-mat/0502054 v1    02 Feb 2005

# Introduction

Molecular dynamics (MD) is computer simulation technique, where the time evolution of a set of interacting particles or atoms is followed by integrating their equations of motion according to the laws of classical mechanics, determined by Newton's law. When particles interact by a pairwise potential only, the thermodynamic ensemble for MD simulations is the microcanonical NVE ensemble. In this case MD simulations are performed in a closed system with volume *V*, containing *N* particles, and the total energy *E* is conserved. Nevertheless, in many cases one wants to modify the equations of motion in such a way that the simulation is performed in an NVT ensemble, where *T* is the absolute thermodynamic temperature, therefore a so-called ''thermostat'' is applied to the system.

There are different types of thermostats for MD simulations in the literature[1-4], each one has advantages and disadvantages. All thermostats give a good temperature control, but they differ in the way how they drive the system to equilibrium. Some correctly represent the hydrodynamic and diffusion behaviour of the system (while having different values of transport coefficients, which are thermostat dependent), while other thermostats drive the system through phase space via some non-physical path, where hydrodynamics is not correctly represented. Here we will outline some of most popular thermostats used in the MD simulations as well their advantages and disadvantages.

One of the most common and simple to implement is the Berendsen thermostat.[1] The idea behind it is to re-scale particle velocities after each time step, so that the desired thermodynamic temperature $T = \Sigma_i m_i v_i^2/3Nk$ is kept constant. All particle velocities are scaled up by the same factor $\lambda = \left[1 + \Delta t / \tau \left(T/T_0 - 1\right)\right]^{1/2}$, where $T_0$ is the desired temperature and $\Delta t$ is the integration time step. Here, $\tau$ is the so-called "rise time" of the thermostat, describing the coupling strength of the system with a hypothetical head bath. The larger the "rise time", the weaker the coupling, i.e. the longer is the period of temperature fluctuations around $T_0$. The Berendsen thermostat is global, since the rescale factor depends on the momentary value of the temperature, which depends on the velocities of all particles in the system. The main disadvantage of this thermostat is that it is not Galilean invariant, and since momentum is not locally conserved, this method does not conserve hydrodynamics.

The Nosé-Hoover thermostat[2-3] is another common choice for canonical MD simulations. The basic idea behind this is to introduce a new internal degree of freedom into Hamiltonian of the system *H*, representing the thermostat coupling. This in turn modifies the equations of motion and introduces one extra equation for the thermostat variable α, which has to be integrated together with the other equations:

$$m_i \mathbf{dr}_i / dt = \mathbf{p}_i, \quad \frac{d\mathbf{p}_i}{dt} = -\nabla_i U - \alpha \mathbf{p}_i,$$

$$\frac{d\alpha}{dt} = \frac{1}{t_s}(T - T_0) \qquad (1)$$

where $\mathbf{p}_i$ is the momentum of particle $i$ with mass $m_i$ at position $\mathbf{r}_i$, $T$ is the momentary value of the system temperature as defined above, $U(\mathbf{r}_i)$ is the potential energy of the system, $t_s$ is the thermostat coupling parameter which controls energy transfer back and forth from the thermostat. In the last equation, α acts as an effective friction parameter. MD simulations in this case are actually performed in micro-canonical $NVE'$, with modified Hamiltonian $H'$ (with energy $E'$). Nevertheless the thermodynamic averages in this $NVE'$ ensemble are equivalent to an average in the canonical NVT ensemble for the original Hamiltonian $H$, but with rescaled particle momentum. This thermostat is global since the momentary value of the temperature is based on a global definition. Obviously, the Nosé-Hoover thermostat is also non-Galilean invariant. Very important for both Berendsen and Nosé-Hoover thermostats, is that the calculations are performed in a reference frame in which the centre of mass of the system is at rest. This is the only frame where velocity rescaling and particle friction relative to the co-ordinate frame preserves total momentum. This restriction also means that the thermostat is not suitable when external forces (like external pressure or gravity) are acting on the system and accelerating the centre of mass. The thermostat effectively brings an additional external friction force, which means that hydrodynamics is artificial. The same holds for total angular momentum, if it is non-zero both Berendsen and Nosé-Hoover will not conserve it as the thermostat introduces an external friction torque. Generally, this thermostat does not conserve local (angular) momentum. These drawbacks are due to the fact that effective thermostating forces are non-central and non-pairwise additive. These disadvantages are removed in the pairwise noise and friction thermostat that is implemented in Dissipative Particle Dynamics (DPD).[4-8]

In the DPD method[4] two pairwise additive and central forces are introduced: a dissipative force $\mathbf{F}_{ij}^D$, and a random force $\mathbf{F}_{ij}^R$. These forces are acting together with conservative force, $\mathbf{F}_{ij}^C$, so that the total force acting on particle $i$ is: $\mathbf{F}_i = \Sigma_j \mathbf{F}_{ij}^C + \mathbf{F}_{ij}^D + \mathbf{F}_{ij}^R$. The sum runs over all neighbouring particles within a distance $r_c$. All forces depend on co-ordinate differences. The random and drag forces are given by $\mathbf{F}_{ij}^R = \xi_{ij} \sigma\, w(r_{ij}/r_c)\, \mathbf{e}_{ij}$, and $\mathbf{F}_{ij}^D = -\frac{1}{2}[\sigma\, w(r_{ij}/r_c)]^2/kT\, (\mathbf{v}_{ij}\cdot\mathbf{e}_{ij})\, \mathbf{e}_{ij}$, where $r_c$ is the cutoff radius, $\xi_{ij}$ are uncorrelated random variables with zero mean and variance 1, and σ is the noise amplitude. The weight function is usually taken as $w(r) = (1-r)$ for $r < 1$ and $w = 0$ for $r > 1$. These forces are tuned in such a manner so that the system evolution in phase space is governed by the Liouville equation, so that it obeys the fluctuation-dissipation theorem.[5] By construction the DPD thermostat is local and Galilean invariant, and therefore preserves hydrodynamics. The main disadvantage of DPD is that the resulting stochastic equations of motion are difficult to integrate self-consistently.[6] Self-consistent integration requires several force calculations per time step, which decreases computational efficiency. An implementation that is not self-consistent[7] (single force calculation per time step), leads to some artefacts. One of these is that the pair correlation function of the ideal gas deviates from one,[6] which means that non-physical interactions are present between particles. The second disadvantage of DPD is that it simulates fluids with comparable diffusion coefficient $D$ and kinematic viscosity $\nu = \eta/\rho$, resulting in Schmidt numbers Sc = $\nu/D$ close to one.[7] Depending on the application this could be a disadvantage, since for most common liquids the Schmidt number is of order of $10^3$, but for diffusion-limited problems this can be viewed as an advantage.

Usually the DPD thermostat is combined with a soft sphere potential: $U_{ij} = ½ a_{ij}(1-r/r_c)^2$ which is the simplest possible potential with cutoff at $r_c$, with a continuous interaction force. Groot and Warren[7] demonstrated relations between the strength of interaction strength $a_{ij}$, and physical observables like liquid compressibility and Flory-Huggins solubility parameters. It is important to emphasize though, that the DPD thermostat is independent of the conservative forces between the particles. It can be used as general purpose MD thermostat with any interaction potential.[7,8] Recently it has been used successfully in combination with Lennard–Jones type of interaction potential.[8]

Yet another completely different thermostat is the Andersen thermostat,[9] which implements a Monte Carlo scheme to sample the equilibrium velocity distribution. The velocity of a randomly chosen particle is replaced by a velocity drawn from a Maxwell distribution. This thermostat is local by nature, but it does not preserve hydrodynamics.

Recently Lowe[10] proposed a generalisation of this Andersen thermostat, which does preserve hydrodynamics and enhances the fluid viscosity. The idea behind this is to change the velocities of a pair of particles, rather than acting on single particles. To this end, the relative velocity is projected on the line connecting their centres, and this value is replaced by a value drawn from a Maxwell distribution. One might interpret this process as the exchange of a virtual particle between the two real particles. It is the momentum carried by these virtual particles that causes viscosity increase. This thermostat is described in more details in the next section. By construction, the Lowe-Andersen thermostat is Galilean invariant, local and preserves hydrodynamics. The viscosity of the fluid sampled with this thermostat is linearly proportional to the exchange frequency, which in turn determines thermostating efficiency. In order to have good thermostating, a relatively high exchange frequency must be used, which in turn leads to very viscous fluids. In some cases this can be disadvantage.

Recently, a generalisation of the Lowe-Andersen thermostat was proposed by Peters.[11] This generalisation effectively leads to a DPD thermostat at zero integration time step and generalised Lowe-Andersen thermostat for finite time steps, allowing the integration of equation of motion and thermostating to be performed in a single time step with a single force route calculation, in a self-consistent manner.

## Local Nosé-Hoover thermostat

The basic idea behind our new thermostat is to combine two thermostats coupled in parallel. The first is a thermostat *similar* to Nosé-Hoover (NHT),[2-3] but which is Galilean invariant and acts on pairs of particles, rather then on single particles. The second is the Lowe-Andersen thermostat (LAT),[10] which is a pairwise analogue of the Andersen thermostat.[9] For each particle pair we choose between NHT and LAT with probability $P = \Gamma\Delta t$, where $\Delta t$ is the integration time step, and where $\Gamma$ is the Lowe-Andersen[10] exchange frequency. Using this approach, we can combine the best of both worlds and gain flexibility, allowing simulations in a broad range of Schmidt numbers, combined with superior temperature control and proper hydrodynamics.

The pairwise analogue of the Nosé-Hoover thermostat is implemented by applying a thermostating force acting on pairs of particles *i* and *j* within a cutoff distance $r_c$.

$$\mathbf{F}_{ij} = -\alpha \psi(r_{ij}/r_c)(1 - T/T_0)\left((\mathbf{v}_i - \mathbf{v}_j) \cdot \mathbf{e}_{ij}\right)\mathbf{e}_{ij}, \qquad (2)$$

where α is a thermostat coupling parameter, $\psi$ is a smearing function chosen such that $4\pi \int_0^1 \psi(r) r^2 dr = 1$ and $\psi(r) = 0$ for $r \geq 1$, $r_{ij} = |\mathbf{r}_{ij}|$ is the distance between particle *i* and *j*, $r_c$ is cutoff radius. $T$ is the momentary value of the system temperature to be defined below, $T_0$ is temperature set value for simulation, $\mathbf{v}_i$ is the velocity of particle *i* and $\mathbf{e}_{ij} = \mathbf{r}_{ij}/|\mathbf{r}_{ij}|$ is a unit vector in the direction of $\mathbf{r}_{ij}$. The coupling parameter α is constant during the simulation, unlike in the Nosé-Hoover thermostat. The advantage of this thermostat is that it conserves both total linear and angular momentum in the system, which is a necessary condition for restoring proper hydrodynamic behaviour. A similar type of thermostating force was independently suggested recently by Phares *et al*,[12] for implementing molecular internal degrees of freedom in energy conserving molecular dynamic simulations. The main difference between their approach and ours is in the temperature dependent pre-factor. In the implementation by Phares *et al*.[12] it depends on the difference between total temperature and mean local internal temperature of particles *i* and *j*, while in our case the pre-factor depends on the difference between momentary and desired temperature of the system. In both cases the thermostating force vanishes at equilibrium, when $T = T_0$. When the temperature differs from equilibrium, the thermostating force given by Eq. (2) either dissipates energy ($T > T_0$) or puts energy back into the system ($T < T_0$). The force also fluctuates due to the fluctuations in mutual particle velocities. So this force serves as both dissipative and random force, but it does so without the use of random numbers. Note that this novel thermostat is therefore completely deterministic, like the Nosé-Hoover and Berendsen thermostats, but unlike Monte Carlo schemes or DPD.

In order to guarantee Galilean invariance, the momentary value of the system temperature should be defined with respect of some relative velocity. The standard reference frame, where thermodynamic and statistical definitions of the temperature coincide, is the frame connected to the centre of the mass of the system. In any other reference frame, moving with a constant velocity *v*, the statistical definition of temperature, i.e. the mean kinetic energy of the system, differs from the thermodynamic temperature. Also, in such a reference frame the velocities of different particles are no longer statistically independent.

Since a fixed frame of reference is unsuitable to define the temperature when hydrodynamic flows are present, we introduce a co-moving frame for each particle. Within this frame the local mean square velocity of the neighbours within a cutoff radius can be determined. Taking the average over all such determined local mean square velocities, we arrive at the temperature

$$kT = \frac{\sum_{i>j} \zeta(r_{ij}/r_c) M_{ij} (\mathbf{v}_i - \mathbf{v}_j)^2}{3 \sum_{i>j} \zeta(r_{ij}/r_c)}. \tag{3}$$

where $\zeta(r)$ is a smearing function for the temperature, chosen such that $\zeta = 0$ for $r > 1$. The cutoff value $r_c$ in the force (Eq. 2) and in the temperature (Eq. 3) could be different in principle. $M_{ij} = m_i m_j / (m_i + m_j)$ is the reduced mass of particles $i$ and $j$. Since at equilibrium the velocities and co-ordinates are independent, it is straightforward to prove that above temperature definition coincides with the equilibrium thermodynamic temperature.

An alternative definition for the temperature follows from defining the co-moving frames for each particle such that the centre of mass of the set of particles within a distance $r_c$ is at rest, rather than the particle itself. This leads to a slightly more complicated definition of the temperature, which in practical simulations turned out to be not significantly different from the temperature defined above. For simplicity we therefore restrict to the above definition.

The fact that the momentary value of the temperature (Eq. 3) is calculated based on all particle velocities and co-ordinates of the system leads to a global thermostat. The Nosé-Hoover thermostat is also global, but this restores hydrodynamics since it is based on a Hamiltonian formalism, while in our case we have a global thermostat which is not based on Hamiltonian. Therefore it is difficult to prove from first principles that the hydrodynamic behaviour of the system is properly conserved. The global nature of this thermostat might be important for hydrodynamics. To determine what happens on a local level (the local thermostating force), global information about particle velocities and their positions is required. In practical terms, however, we expect the method to fail only at very high shear rates, or very inhomogeneous turbulence, since the acceleration of any liquid element equals the sum of forces over its boundaries. This is the defining condition for the Navier-Stokes equation.

For large and dense systems one can use a local temperature definition as well, to implement the thermostat. For instance one can use a local definition of the temperature:

$$kT_i = \frac{\sum_{j} \zeta(r_{ij}/r_c) M_{ij} (\mathbf{v}_i - \mathbf{v}_j)^2}{3 \sum_{j} \zeta(r_{ij}/r_c)}. \tag{4}$$

Provided that $T_i$ does not vary by much (e.g. at high density simulations, or large values of $r_c$ in the temperature smearing function), the thermostating force then can be modified so that it contains only local particle temperatures:[12]

$$\mathbf{F}_{ij} = -\alpha \psi(r_{ij}/r_c) \left[1 - \tfrac{1}{2}(T_i + T_j)/T_0\right] \left((\mathbf{v}_i - \mathbf{v}_j) \cdot \mathbf{e}_{ij}\right) \mathbf{e}_{ij} \tag{5}$$

MD simulations are usually performed with a relatively small number of particles at relatively low particle density. This means that the local temperature defined with Eq. (4) might not be very efficient in terms of temperature control, due to a large variance in local temperature around its mean value. To test this point we performed simulations with a local definition of the temperature (Eqs. 4 and 5). These indeed show a poorer temperature control compared with globally defined temperature (Eqs. 2 and 3), and the fluctuations around $T_0$ are an order of magnitude larger.

A drawback of this variation to the Nosé-Hoover thermostat is that if one starts at zero temperature in a perfectly arranged conformation, all particles will remain at rest and the system will not reach the equilibrium temperature. Though this is a pathological case, it should be kept in mind. If the thermostat is combined with another one that does disturb the initial conformation, this consequence would be circumvented. For the case of the ideal gas, when the initial temperature equals the desired temperature, the thermostat will not perform any work irrespective of the initial velocity distribution. If the initial temperature is slightly off then thermostat will temperate the system and reproduce a Maxwell distribution due to the random nature of the thermostating forces. As a consequence, the diffusion coefficient of ideal gas is divergent since particles move along straight trajectories, and viscosity is zero. The complete lack of thermostat induced interaction between ideal gas particles at the desired temperature guarantees a flat pair correlation function.

## Lowe-Andersen thermostat

As mentioned in the introduction, the Lowe-Andersen thermostat[10] is a pairwise implementation of the Andersen thermostat.[9] In this method a pair of particles that are separated less than $r_c$ is selected to exchange momentum with frequency $\Gamma$. Thus, the mutual velocity along the line between their centres is replaced by a value taken from the Maxwell distribution. The replacement is done such that the total momentum of the particle pair is conserved. The method consists of transforming the particle velocities according to:

$$\mathbf{v}_i := \mathbf{v}_i + M_{ij}\left(\xi_{ij}\sqrt{kT_0/M_{ij}} - (\mathbf{v}_i - \mathbf{v}_j)\cdot\mathbf{e}_{ij}\right)\mathbf{e}_{ij}/m_i$$
$$\mathbf{v}_j := \mathbf{v}_j - M_{ij}\left(\xi_{ij}\sqrt{kT_0/M_{ij}} - (\mathbf{v}_i - \mathbf{v}_j)\cdot\mathbf{e}_{ij}\right)\mathbf{e}_{ij}/m_j \qquad (6)$$

where $\xi_{ij}$ is a Gaussian random variable, and $\langle\xi_{ij}(t)\xi_{km}(t')\rangle = (\delta_{ik}\delta_{jm} + \delta_{im}\delta_{jk})\delta(t-t')$. Again the effective forces arising at each time step are pairwise additive and central, so that total linear and angular momentum of the system are conserved.

In the Lowe-Andersen thermostat (LAT), good thermostating efficiency requires a high collision frequency. This leads to viscous liquids with a low diffusion coefficient, which could be undesirable in some cases. To avoid this, we have combined local Nosé-Hoover with the Lowe-Andersen thermostat in parallel, so that each particle pair is either thermostated by local Nosé-Hoover or via Lowe-Andersen. Thus we are certain that the system will be thermostated whatever the collision frequency. As we will show in the next section this allows simulation of fluids and gases, where the Schmidt number can be varied by several orders of magnitude.

## Numerical implementation

The numerical block scheme for each time step, based on the Galilean invariant Nosé-Hoover-Lowe-Andersen thermostat (NHLAT) is illustrated in Fig. 1. The scheme is based on the velocity Verlet algorithm[13] for integrating the equations of motion. The Nosé-Hoover-like thermostat (NHT) is implemented within the Verlet integration scheme. In the force calculation route, we use estimated values of particle velocity in the intermediate (half integer) time step to calculate the thermostating force, while the momentary temperature is calculated from the velocities in the previous (integer) time step. This thermostat could be implemented separately from the Verlet integration in a consecutive step, but implementing it here increases computational efficiency, since otherwise the nearest neighbour list must be sampled twice. Simulations with separate NHT did not show any significant difference when compared with the algorithm depicted in Fig. 1, which motivates our choice.

In the same force calculation route where NHT is implemented, we build a nearest neighbour list, which is used later on to implement LAT, after finishing the velocity Verlet integration. For the sake of computational efficiency this list stores not only the indexes of particles close to particle $i$, but also the vectors $\mathbf{r}_{ij}$. When the nearest neighbour list is built, the implementation of LAT is straightforward. It is described in the last block diagram in Fig. 1.

When the momentum exchange frequency in LAT is zero, the resulting NHT thermostat is computationally more efficient than standard DPD,[7] since no random numbers are generated. This saves about 3% of computational resource per time step. NHLAT further allows integration with a 50% larger time step, while the system has a higher diffusion coefficient than in DPD. All in all, NHLAT at $P = 0$ is about 70% more computationally efficient than standard DPD.[7] As compared to self-consistent DPD,[6] the thermostat with $P = 0$ is between two and three times faster, while providing similar control over the equilibrium system properties. When P > 0, the computational efficiency of the thermostat is decreased due to the consecutive loop over the selected list of particle pairs (see last part of diagram in Fig. 1). In the worst case, $P = 1$ (LAT only, which corresponds to the highest attainable viscosity) the computational efficiency of the thermostat is about factor of two lower than for standard DPD[7] and is comparable to self-consistent DPD[6]. For realistic values of $P \sim 0.2$-$0.5$, NHLAT is about 10-20% more efficient than standard DPD, but offers similar control over equilibrium system properties as self-consistent DPD.

## Simulation results and discussion

All the simulation results are presented in units where $kT = 1$, $r_c = 1$, and $m = 1$. The masses of all particles are the same. Most of simulations were performed in a box of size 20×20×20 using periodic boundary conditions, with 24000 particles which corresponds to particle density $\rho$ = 3, unless stated otherwise. The time step was $\Delta t = 0.05$. Our DPD code uses velocity Verlet DPD integration, proposed by Groot *et al.*,[7] which differs from the self-consistent integration procedure by Pagonabarraga *et al.*[6] The reason for this is computational efficiency, though it compromises some equilibrium properties like pair correlation distribution of ideal gas (see Fig. 2).

Most of the thermodynamic averages presented here are based on 50,000 time steps simulation, after 10,000 time steps of pre-equilibration run. For the case of DPD simulation, the dissipation coefficient was set to[7] $\gamma = 4.5$ or $\sigma = 3$. For the NHT thermostat we chose $\psi(r_{ij}) = 3(1-r_{ij})/\pi$ and $\alpha = 0.3\pi/(\rho \Delta t)$ (see Eq. 2). Note that, since $\rho = 3$ and $\Delta t$ drops out in the time integration, the velocities at every time step are adjusted by $-0.3(1-r_{ij})(1-T/T_0)(\mathbf{v}_{ij}\cdot\mathbf{e}_{ij})\mathbf{e}_{ij}$. We have performed simulations with three different temperature smearing functions, $\zeta(r_{ij})=1$, $\zeta(r_{ij})=1-r_{ij}$ and $\zeta(r_{ij})=(1-r_{ij})^2/2$, all subject to $\zeta(r_{ij}) = 0$ for $r_{ij} > r_c$ (see Eq. (3)). Since we did not observe any significant differences between these three cases, the results reported here are calculated using the simplest possible choice, $\zeta(r_{ij})=1$.

For the case of interactive fluids the simulations were performed using the DPD soft sphere interactive potential, though DPD and NHLAT are not restricted to this simple force field. In principle both DPD[8] and NHLAT are applicable to any force field providing constant temperature MD simulation, which we believe will conserve hydrodynamics as well.

First we check how NHLAT preserves the equilibrium properties of the system in comparison with standard DPD. In Fig. 2 we show the radial pair correlation function of non-interacting particles (ideal gas) simulated with NHLAT and DPD thermostats. The deviations from g(r) = 1 in DPD are up to 10%, while for the case of our new thermostat they are less then 0.5%. The deviation from g(r) = 1 in DPD is caused by a lack of self-consistency in the integration procedure used, as discussed by Pagonabarraga *et al.*[6] This deviation indicates the presence of a non-physical interaction between the ideal gas particles, induced by the numerical algorithm. For NHLAT such a deviation is not observed, which means that the thermostat and its numerical implementation are free from unphysical artefacts. We also simulated a liquid with the soft repulsive DPD potential at $a = 25$, and compared NHLAT to DPD. The maximum deviation between all pair correlations simulated by DPD and by NHLAT at P = 0, 0.2 and 1, is $10^{-4}$. Hence there is practically no difference between the simulations when conservative forces play a role.

As an additional check of NHLAT we have also sampled the relative velocity distribution, which is presented in Fig. 3a. We compare it with the expected Maxwell distribution without any adjustable parameters, $P(x) = 3\sqrt{3x/(2\pi)}\exp(-3x/2)$, where $x = M_{ij}(\mathbf{v}_i-\mathbf{v}_j)^2/kT$. There is obviously an excellent agreement between expected and simulated distribution. In our NHLAT simulations at $\Delta t$ = 0.05 in a 10×10×10 box and density $\rho$=3, the mean temperature deviation from $T_0$ is about 0.004% for $P$ = 0, and –0.04% when $P$ = 0.2. The typical temperature error in DPD ($\gamma = 4.5$) is about 0.4% for the same system size and time step. Even at step size $\Delta t$ = 0.1 the temperature error in NHLAT is only 1.4%. To have the same temperature accuracy as in Groot-Warren DPD[7] at step size $\Delta t$ = 0.06 (1%), this method allows time steps of $\Delta t$ = 0.09, i.e. 50% larger steps. Also the fluctuation amplitude about the mean value is smaller in $P$ = 0 NHLAT than in standard DPD. We find a Gaussian distribution with a standard deviation of 0.8% (for 3000 particles) in NHLAT at $P = 0$. At $P = 0.2$ it is 1.4%, which is comparable to the 1.5% variance obtained in DPD. The mean temperature does not depend on system size but the fluctuation amplitude decreases as $1/\sqrt{N}$. The efficacy by which the temperature is kept constant is demonstrated in Fig 3b, which shows the temperature autocorrelation for DPD and NHLAT at $P = 0$. Clearly, NHLAT has a much faster temperature control than standard DPD. Although there are quite a few integration algorithms for DPD, for soft sphere interactions most have comparable errors in their temperature control.[14] The only exception is the integration scheme SC-Th,[14] which obtains the correct temperature by dynamically fine tuning the friction, but the consequence of this is that transport coefficients become strongly time step dependent. We do not find such time step dependence in the diffusion constant for NHLAT.

Additionally we have compared DPD and NHLAT in predicting surfactant phase behaviour, based on dissolution of a concentrated surfactant phase in water.[15] At the same value of interaction parameters and system composition, DPD and NHLAT are leading to a similar phase structures. Since NHLAT allows bigger time steps and the diffusion coefficient is higher, the equilibrium phase is reached about 1.5 times faster when compared to DPD.

Having established by simulation that NHLAT generates the proper equilibrium, we now focus on the transport properties of the system. The diffusion coefficient is determined via the time dependent displacement. For simulations with box size $10 \times 10 \times 10$ we find that the Langevin equation applies, which is solved by $\langle (r(t)-r(0))^2 \rangle / 6t = D\,[1 - mD/(kTt) + mD/(kTt)\exp(-kTt/mD)]$. Since this equation applies for all times t, this equation allows us to fit the diffusion constant on the basis of all data points rather than taking the large time limit.

In Fig. 4 we show the dimensionless diffusion coefficient as a function of interaction parameter obtained using NHLAT with $P=0$. In the limit of small repulsion parameters, we find a divergent diffusion constant $D \sim (15.9 \pm 0.4) a^{-1.75 \pm 0.02}$. Though this might look striking at first, this actually is the correct physical behaviour of a non-interacting gas. For the truly ideal gas, momentum is conserved per particle, leading to ballistic motion and hence to a divergent diffusion constant. NHLAT ($P=0$) does not destroy this momentum conservation per particle, and hence correctly reproduces a divergent diffusion constant. Nevertheless, the system is kept at the desired temperature with very high precision, even when external shear or sine forces are applied in the system. For a non-ideal fluid, the interaction between the particles decreases the diffusion constant, and we find a crossover to a different power-law. This indicates liquid-like behaviour. When the repulsion is increased above roughly $a \sim 550$ a weak solid is formed, where the diffusion constant vanishes. A good fit to the observed diffusion behaviour is $D = 15.9 a^{-1.75} + 0.96 a^{-0.33} \exp(-a/203)$, see Fig. 4. In DPD at $\sigma = 3$, the diffusion constant is determined by the repulsive interaction for $a > 10$ and the results coincide with NHLAT, but friction slows down diffusion significantly for $a < 10$. This implies that for strongly interacting systems the transport coefficients are determined by the conservative force field, whereas for weakly interacting systems the thermostat may contribute significantly to viscosity and diffusion constant. For instance, in DPD with noise $\sigma = 3$, diffusion is completely determined by the repulsion for $a > 15$.

When the collision probability $P$ is different from zero the ideal gas has a finite diffusion coefficient: for $P=0.2$ the diffusion coefficient is $D = 0.17$, while for DPD the ideal gas diffusion coefficient is $D=0.71$. A very high diffusion coefficient at $P=0$ makes NHLAT very attractive and efficient to simulate weakly interacting gases. When combined with electrostatic interactions[16] this will allow realistic plasma simulations.

Viscosity was estimated by applying a sinusoidal force in the x-direction over a given period of time $t \in [0, t_0]$ and studying the response. This sine force $f_x = A_f \theta(t_0 - t) \sin(ky)$ is periodic in the y-direction, where $k = 2\pi/L$ and $L$ is the box length in the y-direction, while $\theta(t_0 - t)$ is a step function vanishing at $t > t_0$. The Navier-Stokes equation for a system with such an external force is given by

$$\frac{\partial v_x}{\partial t} = A_f \theta(t_0 - t) \sin(ky) + \nu \frac{\partial^2 v_x}{\partial y^2}. \quad \text{For} \quad v_x(y,0) = 0 \quad \text{this has the solution}$$

$$v_x(y,t) = \frac{A_f}{\nu k^2} \sin(ky) \left\{ \left(1 - e^{-\nu k^2 t}\right) + \left(e^{-\nu k^2 (t-t_0)} - 1\right) \theta(t - t_0) \right\}.$$ The solution implies that if the kinematic viscosity is independent of the wave vector $k$, we should get a data collapse if we plot $2 < v_x(y,\tau) \sin(ky) > k^2$ as a function of $\tau = k^2 t$ at constant $A_f$ and $\tau_0 = k^2 t_0$. In Fig. 5 we show the result from three such simulations. Indeed the simulation data practically collapse to one curve, with a response as predicted from the theory. From the best fit we thus obtain $\nu = \eta/\rho = 0.257 \pm 0.003$. When no external forces are applied, viscosity can be estimated from the stress auto-correlation function[10] using the Green-Kubo relation. From this analysis we obtain the independent value $\nu = \eta/\rho = 0.264 \pm 0.004$, which is very close to the value obtained from dynamical simulations. The fact that we are able to fit all response curves with a single parameter (kinematic viscosity) solution of the Navier-Stokes equation, and that two independent estimations of viscosity give very close results, shows that the proposed novel thermostat properly restores the hydrodynamic behaviour of the system.

To finish the discussion, we have performed NHLAT simulations at fixed interaction constant $a=25$, while we have varied the probability $P = \Gamma \Delta t$ at constant time step $\Delta t = 0.05$. In Fig. 6 we show the dependence of diffusion coefficient and estimated fluid viscosity as a function of collision probability $P$. For simulations with $P>0$ fluid viscosity is proportional to the collision probability $P$, while the diffusion coefficient is inversely proportional to it.[10] This means that the Schmidt number Sc = $\nu$/D is proportional to the square of the collision probability. All these scaling relations are confirmed by the simulation data presented in Fig. 6. By changing the collision probability, the Schmidt number can be varied by more than two orders of magnitude, whereas in standard DPD it is always close to 1. By decreasing $\Delta t$, the Schmidt number may take any desired value. Since the system is always thermostated, the simulation can be performed at any value of collision probability (even zero), which is a significant advantage over the LAT, which fails at low values of $P$, where the systems is very weakly thermostated.

## Summary and Conclusions

In the present article we propose a novel thermostat for molecular dynamic simulations, which is a combination of pairwise Nosé-Hoover[2-3] and Lowe-Andersen[9-10] thermostats. The Nosé-Hoover-like thermostat is implemented in a Galilean invariant way by the introduction of a pairwise force, which is proportional to the projected relative velocity and to the difference between required and actual temperature in the system. The Lowe-Andersen thermostat replaces the projected relative velocity of a pair of particles by a value obtained from a Maxwell distribution. For each particle pair, NHT and LAT are chosen with probability $P = \Gamma \Delta t$, which assures thermostating at any value of the *P*. The case of *P* = 0 corresponds to the pairwise variation of the Nosé-Hoover thermostat, while the case *P* = 1 gives a Lowe-Andersen thermostat. When NHT dominates (low *P*) the simulated liquid has a high diffusion coefficient and low viscosity, and when LAT dominates the reverse is true.

The proposed Nosé-Hoover-Lowe-Andersen thermostat (NHLAT) is Galilean invariant and preserves total linear and angular moment of the system, which is due to the fact that the thermostatic forces act between pairs of the particles. Moreover, the force is pairwise additive and central, and the temperature is calculated in a Galilean invariant manner, using relative particles velocities. Though we have no theoretical proof, we expect that above mentioned properties lead to proper hydrodynamic behaviour of the system, since the acceleration of each liquid element is exactly given by the sum of forces over the boundary. Indeed, the kinematic viscosity determined from a force balance is consistent with the result obtained from the dynamic response time. The only possible drawback that we anticipate is that the temperature is obtained as an average over the whole system, which could lead to a subtle non-local interaction.

The NHLAT (*P*=0) ideal gas has a divergent diffusion coefficient and zero viscosity, while for *P*>0 it has finite diffusion coefficient. Thus, by adjusting the probability *P* in NHLAT, the Schmidt number of the system can be varied by several orders of magnitude. The (*P*=0) diffusion constant of the interacting liquid approaches zero at very high repulsion parameters where the system forms a soft solid.

The temperature deviation from the required system temperature with NHLAT is at least an order of magnitude smaller than for standard DPD,[7] while the equilibrium properties of the system are very well reproduced. At the same time NHLAT is computationally more efficient than self-consistent DPD,[6] by offering better temperature control and larger flexibility in terms of adjusting diffusion coefficient and viscosity.

The presented thermostat can be applied in any conventional particle based molecular dynamics simulation, including atomistic force fields. In particular, simulation of transport coefficients of dilute gases and plasmas is possible. Whereas other thermostats will decrease the diffusion constant for dense nearly ideal fluids, this method preserves the divergence of the mean free path at small interaction. Generalisations of the methods are straightforward. For a two-phase system one can choose three exchange different frequencies $P_{11}$, $P_{22}$, and $P_{12}$. Thus, if phases 1 and 2 are mutually insoluble and if $P_{12} > P_{11} \geq P_{22}$, parameter $P_{12}$ will introduce an excess surface viscosity of the resulting interface. With an extra surface active component into the system, this would open up a new class of simulations where both bulk and surface viscosities and diffusion coefficients could be adjusted at will. The method can also be used to generate an NVE ensemble, if in Eq. (2) the thermostated parameter $T/T_0$ is replaced by the actual Hamiltonian divided by the required energy, $H/E$. Thus, numerical errors in the integration algorithm are compensated by a local, pairwise additive force, so that a truly adiabatic system is simulated, including hydrodynamics.

# REFERENCES

[1] H. J. C. Berendsen, J. P. M. Postma, W. F. VanGunsteren, A. DiNola and J. R. Haak., *J. Chem. Phys.* **81,** 3684 (1984).

[2] S. Nosé, *J. Chem. Phys.* **81**, 511 (1984).

[3] W. G. Hoover, *Phys. Rev. A* **31**, 1695 (1985).

[4] J. M. V. A. Koelman and P. J. Hoogerbrugge, *Europhys. Lett.* **21**, 363 (1993).

[5] P. Español and P. Warren, *Europhys. Lett.* **30**, 191 (1995).

[6] I. Pagonabarraga, M. H. J. Hagen and D. Frenkel, *Europhys. Lett.* **42**, 337 (1998).

[7] R. D. Groot and P. B. Warren, *J. Chem. Phys.* **107**, 4423 (1997).

[8] T. Soddemann, B. Dunweg and K. Kremer, *Phys. Rev. E* **68**, 046702 (2003).

[9] H. C. Andersen, *J. Chem. Phys.* **72**, 2384 (1980).

[10] C. P. Lowe, *Europhys. Lett.* **47**, 145 (1999).

[11] E. A. J. F. Peters, *Europhys. Lett.* **66**, 311 (2004).

[12] D. J. Phares and A. R. Srinivasa, *J. Phys. Chem. A* **108**, 6100 (2004).

[13] W. C. Swope, H. C. Andersen, P. H. Berens and K. R. Wilson, *J. Chem. Phys.* **76**, 637 (1982).

[14] I. Vattulainen, M. Karttunen, G. Besold and J. M. Polson, *J. Chem. Phys.* **116**, 3967 (2002).

[15] P. Prinsen, P. B. Warren and M. A. J. Michels, *Phys. Rev. Lett.* **89**, 148302 (2002).

[16] R. D. Groot, *J. Chem. Phys.* **118**, 11265 (2003).

# Figure Captions

FIG 1. Novel Nosé-Hoover-Lowe-Andersen thermostat (NHLAT) - block scheme of single time step implementation based on the velocity Verlet integration algorithm.

FIG 2. Radial pair correlation function of deal gas, a comparison between DPD and Nosé-Hoover-Lowe-Andersen thermostat (NHLAT).

FIG 3a. Distribution of relative velocities in NHLAT and comparison with equilibrium Maxwell distribution, $P(x) = 3\sqrt{3x/2\pi} \exp(-3x/2)$, without any adjustable parameters. All simulations were performed at k$T$=1.

FIG 3b. Time dependence of normalised temperature autocorrelation function for DPD and NHLAT at *P*=0.

FIG 4. Dimensionless diffusion coefficient as a function of interaction parameter *a/kT*, at density $\rho r_c^3 / m = 3$, using NHLAT at *P*=0 and using DPD at σ = 3.

FIG 5. Time evolution of average particle velocity in the direction of applied sine force for different values of wave number *k* (box size). The sine force was applied at t=0 and switched off at $tk^2 = 60$. The force amplitude in all of the cases was $A_f = 0.05$, density was $\rho = 3$ and P=0. The solid line represents a fit of the average of all three simulations to a single parameter solution of the Navier-Stokes equation with kinematic viscosity $\nu = \eta / \rho = 0.257 \pm 0.003$. The two inset graphs show the evolution at the beginning and shortly after force was switched off.

FIG 6. Dependence of NHLAT inverse diffusion coefficient *D* and kinematic viscosity ν as a function of collision probability $P = \Gamma \Delta t$, with Δt = 0.05. The simulations are performed at *a/kT*=25 and $\rho = 3 m / r_c^3$. Viscosity was determined by applying an external sinusoidal force. From linear fits to the data we get $\nu = 0.26 + 2.7P$ and $D^{-1} = 2.2 + 30.4P$. The Schmidt number in this case varies from 0.5 to 100. By decreasing Δt the Schmidt number may take any value higher than this.

# Figures

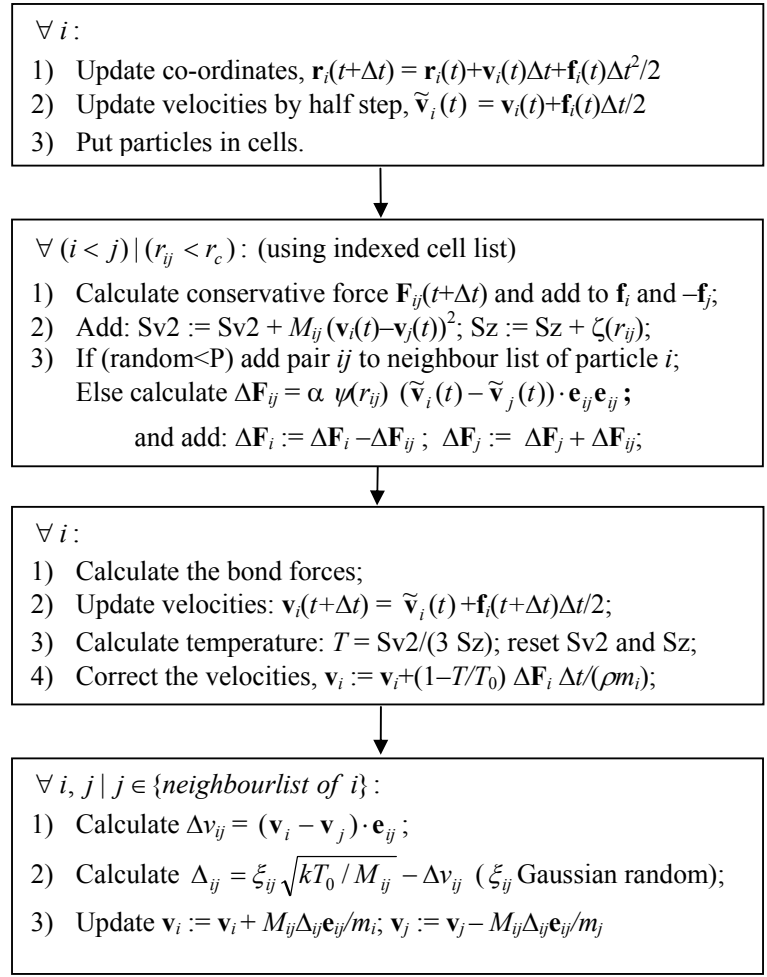

FIGURE 1.

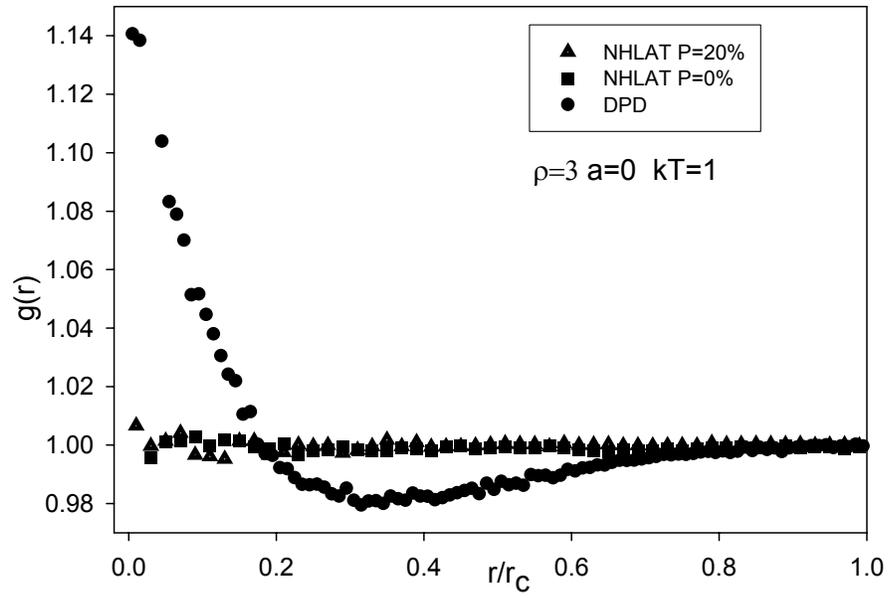

FIGURE 2.

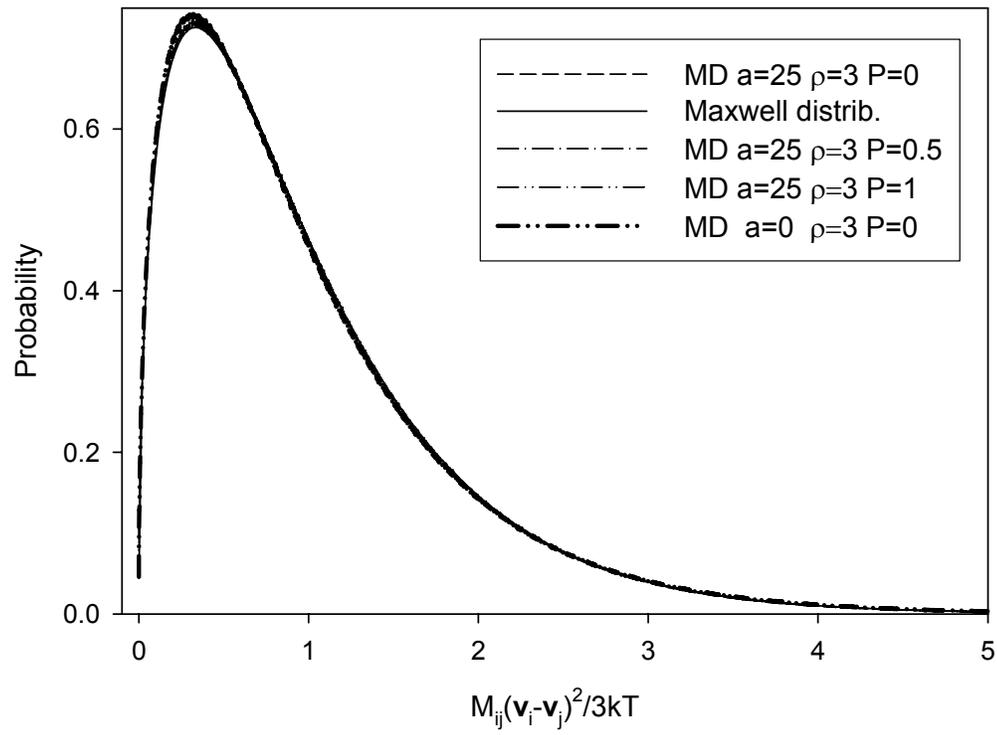

FIGURE 3a.

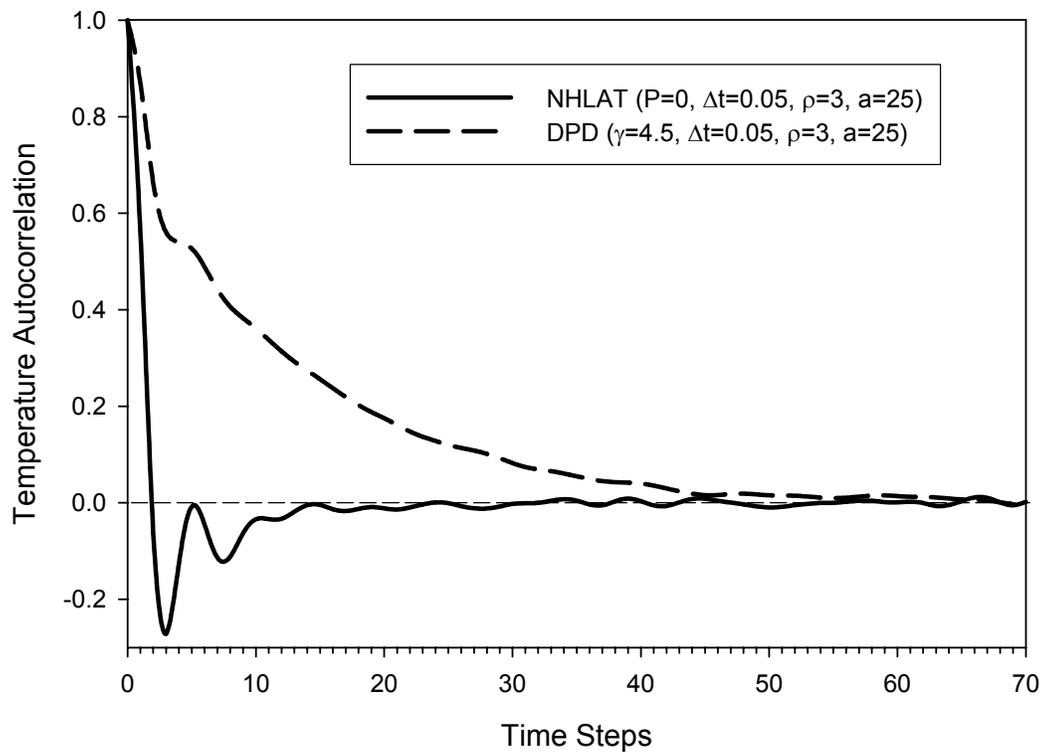

FIGURE 3b.

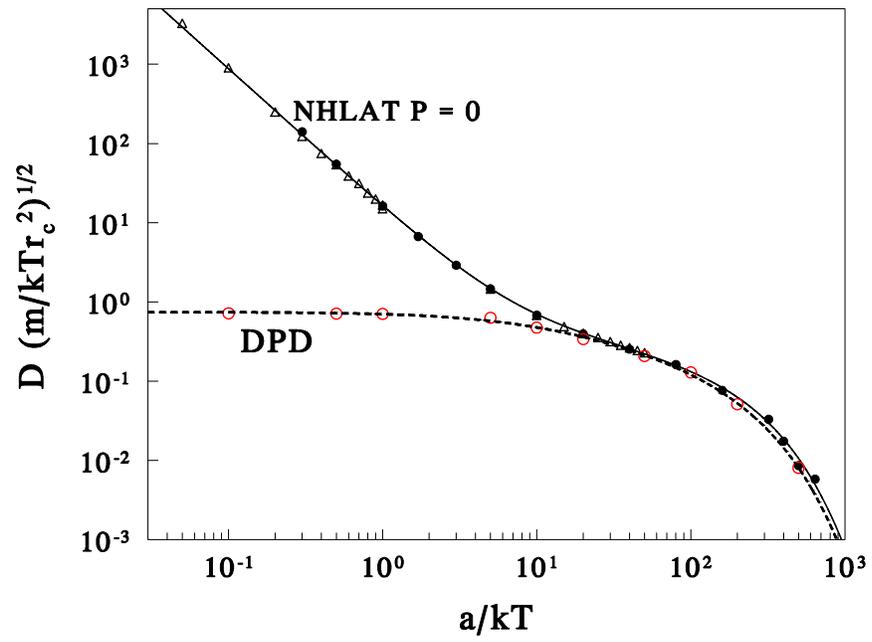

FIGURE 4.

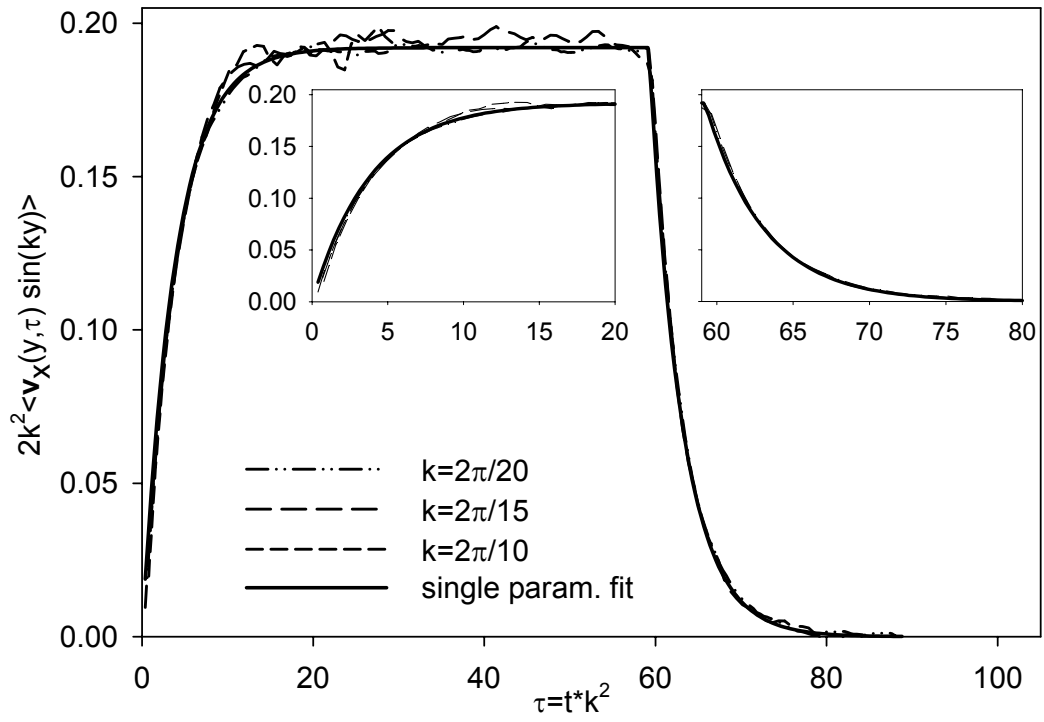

FIGURE 5.

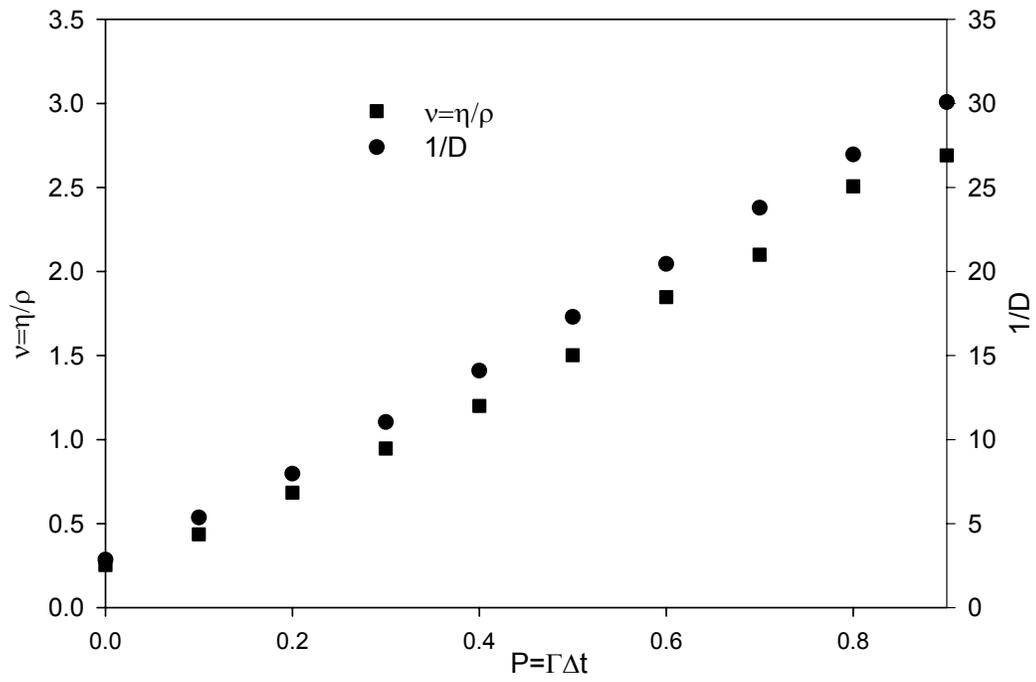

FIGURE 6.